\documentclass[aps,pra,reprint,nofootinbib,superscriptaddress]{revtex4-2}
\usepackage{CJK}
\pdfoutput=1
\usepackage[utf8]{inputenc}
\usepackage[english]{babel}
\usepackage[T1]{fontenc}
\usepackage{amsmath}
\usepackage{svg}
\usepackage{hyperref}
\hypersetup{
    colorlinks=true,
    allcolors=blue,
    }

\usepackage{xcolor}
\usepackage{datetime}
\usepackage{amssymb}  
\usepackage{bm}
\usepackage{parskip}
\usepackage{cancel}
\usepackage{tikz}
\usepackage{lipsum}
\usetikzlibrary{quantikz}
\usepackage[ruled,vlined]{algorithm2e}
\usepackage{orcidlink}

\newcommand{\figref}[1]{\figurename~\ref{#1}}
\def\be{\begin{equation}}
\def\ee{\end{equation}}
\def\GS{\hbox{GS}}
\def\TFD{\hbox{TFD}}
\def\beqa{\begin{eqnarray}}
\def\eeqa{\end{eqnarray}}
\def\btheta{\bm{\theta}}
\def\bphi{\bm{\phi}}
\def\bvartheta{\bm{\vartheta}}
\def\bvarphi{\bm{\varphi}}
\def\blambda{\bm{\lambda}}
\renewcommand{\ket}[1]{\ensuremath{|#1\rangle}\xspace}
\renewcommand{\bra}[1]{\ensuremath{\langle #1|}\xspace}
\renewcommand{\braket}[2]{\ensuremath{\langle #1|#2\rangle}\xspace}

\date{\today}
\begin{document}
\title{Hamiltonian Forging of a Thermofield Double}

\author{Daniel Faílde\orcidlink{0000-0002-7685-4331}}
\thanks{D.F and J.SS contributed equally to this work.}
\email{dfailde@cesga.es}
\affiliation{Galicia Supercomputing Center (CESGA), 15705 Santiago de Compostela, Spain}
\author{Juan Santos-Suárez\orcidlink{0000-0001-9360-2411}}
\thanks{D.F and J.SS contributed equally to this work.}
\email{juansantos.suarez@usc.es}
\affiliation{Departamento de  F\'\i sica de Part\'\i  culas, Universidade de Santiago de Compostela and Instituto Galego de F\'\i sica de Altas Enerx\'\i as (IGFAE), 15782 Santiago de Compostela, Spain}
\author{David A. Herrera-Mart\'i\orcidlink{0000-0003-3606-7998}}
\email{david.herrera-marti@cea.fr}
\affiliation{Universit\'e Grenoble Alpes, CEA List, 38000 Grenoble, France}
\author{Javier Mas\orcidlink{0000-0001-7008-2126}}
\email{javier.mas@usc.es}
\affiliation{Departamento de  F\'\i sica de Part\'\i  culas, Universidade de Santiago de Compostela and Instituto Galego de F\'\i sica de Altas Enerx\'\i as (IGFAE), 15782 Santiago de Compostela, Spain}

\begin{abstract}
We address the variational preparation  of the Thermofield Double as the ground state of a suitably engineered Hamiltonian acting on the doubled Hilbert space. 
Through the use of the {\em Entanglement Forging} ansatz, we propose a solution that involves only circuits of width $N$. We illustrate the method with generic fermionic Hamiltonians. The free fermion case can be solved in closed form, and yields a warm start state for the variational circuits whenever interactions are present. As an important side product, this method returns the  complete energy spectrum and eigenbasis of the system.
\end{abstract}

\maketitle

\section{Introduction}

State preparation is of central importance in quantum computation, and it cannot be disentangled from the algorithmic prescription of a quantum task. When dealing with circuits that simulate quantum thermal processes, both at and out of equilibrium, the initial state that needs to be preparated is, typically, the Gibbs state. Its purification corresponds to the Thermofield Double (TFD) state, and preparing it can be as challenging as finding the ground state of a general Hamiltonian \cite{watrous2008quantum, aharonov2013guest}. Such initial state has emerged as a valuable tool to study the thermal behavior of quantum systems \cite{de2015thermofield, de2017dynamics, tamascelli2019efficient, nusseler2020efficient}. It is also relevant for the simulation of  gravity duals to black holes and wormholes  \cite{Brown:2019hmk,Lata:2021}. 

Variational quantum algorithms have become a practical method to prepare quantum states across different domains in quantum simulation.  In relation to the present context, a number of works have proposed to generate  the TFD state by minimizing the Helmholtz free energy  \cite{wu2019variational, chowdhury2020variational,Sagasti2021}. They differ mainly in the choice of the variational ansatz. All of them, however, face the difficulty of having to compute the von Neumann entropy on the fly, which is not an observable. Some ideas have been proposed to accomplish this task  \cite{wang2021variational,Foldager:2021qyk,consiglio2023variational}.

In this paper, we also address the preparation of the TFD state by variational quantum computation but, instead of the free energy, we propose to minimize the energy of an ad hoc designed Hamiltonian. The idea is inspired by a similar mechanism suggested and studied in  \cite{maldacena2018eternal,Cottrell_2019,Alet_2021} although we will be working in the opposite weakly coupled  limit that perturbs around the free fermion case, where the solution is exact. This solution will act as a {\em warm start} for the variational optimization,  thereby reducing its complexity.

On top of that, we notice that the structure of the TFD itself is very well adapted to the use of the so called \emph{Entanglement Forging} (EF) \cite{eddins2022doubling} formalism where the role of the Schmidt coefficients is played by the Boltzmann weights. This is a simplified version of the circuit {\em cutting/knitting} technique which allows addressing the preparation of certain $2N$ qubit bipartite states, such as the TFD, by running only simpler quantum circuits of half width, hence $N$ qubits. 

An extra bonus is that, by design, the minimization of the variational  circuit yields approximately the unitary matrix that diagonalizes the Hamiltonian of the system. This feature, which can also be found in \cite{consiglio2023variational}, provides  access to the whole spectrum  and is an alternative to standard deflationary approaches \cite{higgott2019variational}, which are based on iteratively adding penalty terms associated to rank-1 projectors.  Alternative approaches can be found in the literature that make use of the Schmidt decomposition ansatz to address similar problems \cite{Benavides-Riveros:2022two,Hong:2023atp, Bravo-Prieto}.

This paper is structured as follows. In Sec.~\ref{sec:TFD} we introduce the main idea and exemplify its performance on a generic fermionic model.  In Sec.~\ref{sec:forging} we present the variational forging ansatz used to actually prepare the TFD state on a quantum computer. Finally, in Sec.~\ref{sec:results} we show the results and conclude in Sec.~\ref{sec:conclusions} with comments.

\section{TFD States as Ground States}\label{sec:TFD}

Given a  Hamiltonian $H$, the associated Gibbs state, $\rho = e^{-\beta H}/Z$, where $Z\equiv \hbox{Tr}\left(e^{-\beta H}\right)$ characterizes equilibrium in the canonical ensemble at temperature $T=1/\beta$. The purification of this state, the TFD, requires to duplicate the number of degrees of freedom of the system  \cite{Takahashi}. We will focus on purifications of the following form
\be
\ket{\hbox{TFD}(\beta)} =\frac{1}{\sqrt{Z}}\sum_i e^{-\beta E_i/2} \ket{E_i}\otimes \ket{E^*_i}, \label{TFD}
\ee
where $\ket{E^*_i}$ stands for the complex conjugate of the energy eigenbasis.\footnote{
In general, there could be an arbitrary anti-unitary operation involved $\ket{\bar E_i} = U\ket{E_i}$, but we will stick to the minimal option.  }
The promise is that there exists a certain Hamiltonian, $H_{tot}(\beta)$, in the enlarged Hilbert space, whose ground state $\ket{\hbox{GS}(\beta)}$ is, with high overlap, the sought-after TFD 
$$
|\braket{\GS(\beta)}{\TFD(\beta)} | \approx 1 \,.
$$

Concrete expressions for $H_{tot}$ have been proposed in the literature, and they range from  exact but impracticable \cite{Cottrell_2019}, to  simple but approximate   \cite{Cottrell_2019,maldacena2018eternal,Alet_2021}.  In  the second case, the proposed Hamiltonians have the generic structure $H_{tot}(\beta) = H_L + H_R +  H_{LR}(\beta)$, where $H_L = H\otimes I$, $H_R = I\otimes H^*$ and, therefore, it only remains to find $H_{LR}$. 
 A very general prescription is that $H_{LR}(\beta)$ is a Hamiltonian that should dominate over $H_L+H_R$ in the low $\beta$ regime $H_{tot}\left(\beta \to 0\right)\approx H_{LR}\left(\beta \to 0\right)$ where $\ket{\GS\left(\beta \to 0\right)}=\ket{\TFD(\beta \to 0)}$  that is, the ground state of $H_{LR}$ should feature maximal entanglement. At the opposite end, i.e. at low temperature, $H_{LR}(\beta \to \infty)  $ should vanish, so that $H_{tot}(\beta \to \infty ) = H_L + H_R$ where, by definition $\ket{\GS(\beta \to \infty)} = \ket{E_0}\otimes \ket{E^*_0}=\ket{\TFD(\beta \to \infty)}$.

Outside these limits,  $H_{LR}(\beta)$ should be carefully adjusted in order  to keep the overlap $|\braket{\hbox{GS}(\beta)}{\hbox{TFD}(\beta)}| $ as close to 1 as possible.  In the cited references,  no controlled scheme is offered whereby one can obtain an estimation of the departure from maximal fidelity.  In the next section, we seek for such a protocol by starting from an exact answer valid for the case of free fermions. The introduction of an interaction degrades this maximal fidelity in a continuous way. We provide a set of rules to still find an excellent answer. 

\subsection{Fermionic models}

We  start by looking at  fermionic  systems and assume to work in a quasi-particle basis in which the free (quadratic) piece has been brought to its diagonal form.\footnote{Diagonalizing a quadratic Hamiltonian is a polynomial  task in resources, and therefore, we will assume this has been performed with classical computation  \cite{Surace2021}. } Hence, the general  case can be written as follows
\begin{eqnarray}
H &=& \sum_{i=1}^N \omega_i a^\dagger_i a_i  + \sum_{ijkl=1}^N U_{ijkl} a_i^\dagger a_j^\dagger a_k a_l + ... \nonumber \\ 
&=& H_2 + H_4 + ... \label{hamil}
\end{eqnarray}
with $\omega_i \in {\mathbb R}$ nonzero real numbers and $N$ the dimension of the Hilbert space.  
To start with,  consider the free-fermion limit $H = H_2$. To find the TFD as a ground state, we double the Hilbert space and  enlarge accordingly the set of operators $a_i \to a_i^L = a_i\otimes I, a_i^R=I\otimes a_i$.  In this free-fermion limit, an exact  solution for
\be
H_{tot}(\beta) = H_L + H_R + H_{LR}, \label{htot}
\ee
exists, and is  given by 
\begin{eqnarray}
H_{L} &=& \sum_i \omega_i a^{L \dagger}_i a^L_i  ~~,~~H_{R} =   \sum_i \omega_i a^{R \dagger}_i a^R_i~~, \label{hLRLR}\\
H_{LR} &=& \sum_i \mu_i(\beta) ( a^{L}_i a^{R}_i - a^{L\dagger}_i a^{R\dagger}_i),  \nonumber
\end{eqnarray}
with
\be
\mu_i (\beta) = \frac{\omega_i}{2\sinh \beta\omega_i/2} \, . \label{muofbeta}
\ee

A proof of this statement can be found in App.~\ref{sec:app_A}. The overlap $|\braket{\hbox{GS}(\beta)}{\hbox{TFD}(\beta)}| = 1$ is maximal,  for all values of $\beta$.\footnote{In fact, $\ket{\hbox{GS}(\beta)}$ is the ground state of a uniparametric family of Hamiltonians. Apart from  $H_{tot}$ in \eqref{htot}, two important members of this family are  also $H_L - H_R$ and  $H_L  + \frac{1}{2}H_{LR}$. For the second, this provides a short proof for the Entanglement Cloning Theorem \cite{Hsieh_2016}. See App.~\ref{sec:app_A} for more information.}

A four-fermion interaction like $H_4$  in \eqref{hamil}  makes the situation depart from the free exact case.   The question now is twofold: (a) how should we modify the interaction Hamiltonian $H_{LR}$ so as to maintain the overlap as close to one as possible?, and (b) how large can $H_4$ be made while staying globally within some tolerance, say, $|\braket{\hbox{GS}(\beta)}{\hbox{TFD}(\beta)}| \geq 0.99$.\footnote{Of course, an implicit assumption in the spirit of any perturbative approach is that, in making $H_4$ large, we do not hit any quantum phase transition. }

For the first question, a minimal answer would be to stick to the same structure as in \eqref{htot} and \eqref{hLRLR}, and only modify the set of numbers  $\omega_i \to \tilde\omega_i = \omega_i +  \delta \omega_i $   which are then plugged into \eqref{muofbeta}  to construct the improved  LR interaction. 
The  shifts $\delta\omega_i$  should be readable  from the structure of the perturbing Hamiltonian $H_4$. An educated guess is to extract them from a mean-field calculation. 

In regard to the second question, the construction now is no longer exact and we expect the overlap to drop below 1 for  intermediate values of $\beta$.\footnote{ Notice that, from \eqref{muofbeta} we have that $H_{tot}(\beta\to\infty) \to H_L + H_R$ whereas $H_{tot}(\beta \to 0) \to H_{LR}$. This makes the overlap 
$|\braket{\hbox{GS}(\beta)}{\hbox{TFD}(\beta)}| =1$ in these two limits $\beta \to 0,\infty$.} The  answer needs a case-by-case analysis. We illustrate it with an example in the following sections.

\subsection{Numerical test: the Hubbard spinless chain}

We will consider  a popular benchmark model, the Hubbard 1D chain. Since we are not interested in magnetization phenomena, we will consider a simplified version with spinless fermions on $N$ sites:
\be
H = \sum_{i=1}^N \epsilon_0  a^\dagger_i a_i  - t \left( \sum_{i=1}^N     a^\dagger_i a_{i+1}  + h.c \right)+ U\sum_{i=1}^N  a_i^\dagger a_i a_{i+1}^\dagger a_{i+1}  \, . \label{Hub_real}
\ee
Here, the quadratic terms encode  the on-site potential and nearest-neighbor interactions. They are weighted by a chemical potential $\epsilon_0$ and hopping amplitude $t$, respectively. The quartic terms introduce nearest neighbor repulsion  parametrized by a constant $U$. To bring this Hamiltonian into the form \eqref{hamil}, the change of basis needed  is just a simple Fourier transform (assuming periodic boundary conditions), after which
\be
H = \sum_{k} \omega_k  a^\dagger_k a_k  + \frac{U}{N}  \sum_{k,p,q}   e^{-i\frac{2\pi}{N}q} \;  a^\dagger_{k+q} a_{k}  a^\dagger_{p-q}  a_{p} \,,
\label{Hub_k}
\ee
where $\omega_{k}=\epsilon_0-2t \cos(2\pi k/N)$, with $k =0,..,N-1$. The interaction  term proportional to $U$ is non-local in momentum. It  gives rise  now to scattering events between two incoming electrons with momentum $k$ and $p$ that exchange momentum $q$.  As previously discussed, the model with only $H_2$ is exactly solvable. Adding $H_4$ makes  the overlap $|\braket{\hbox{GS}(\beta)}{\hbox{TFD}(\beta)}|$  decrease if this quartic interaction introduces a significant contribution that competes with  the diagonal $H_2$. In this example, such contributions arise from scattering terms with net momentum exchange zero, i.e., $q=0$ and $k+q=p$. The non-interacting energies $\omega_k$ are not the best candidates to estimate $H_{LR}$ and have a ground state with optimal overlap with the TFD. 

\begin{figure}[htbp]
\begin{center}
\includegraphics[scale=0.9]{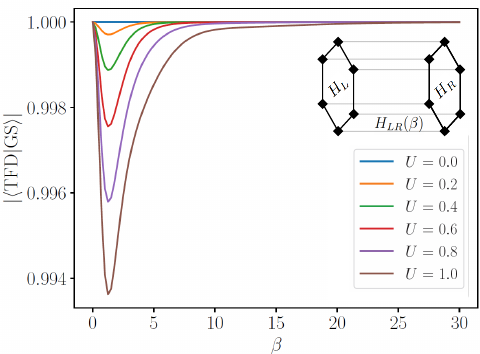}
\caption{$\left|\braket{\hbox{TFD}(\beta)}{GS(\beta)}\right|$ for the Hubbard model in frequency space ($t=1$, $\epsilon_0=0$, $2N=12$) for different values of $U$. For every $U$, mean-field energies $\tilde\omega_k (\epsilon_0,t,U)$ were estimated numerically by minimizing the free energy. The figure shows how the prescription for $H_{LR}$ leads to \hbox{TFD}s with high overlaps with the GS of $H_{tot}$.}
\label{fig:overlap_Hubbard}
\end{center}
\end{figure}

As previously advertised, an educated guess to improve the \textit{efficacy} of the interaction Hamiltonian  $H_{LR}$ involves shifting the non-interacting 
\be
\omega_k \to \tilde \omega_k = \omega_k + \delta \omega_k, \label{wshift}
\ee
appropriately.
A natural candidate for $\delta \omega_k$ comes from the diagonal mean-field contribution of $H_4$ \cite{bruus2004many, Pavarini:819465}, for which $\delta \omega_k  =  2U\rho-2U\alpha \cos{(2\pi k/N)}$.  Here, $\rho = \frac{1}{N}\sum_{k'} \langle a_{k'}^\dagger a_{k'}\rangle$ is the mean-field value of the density and $\alpha = \frac{1}{N}\sum_{k'} \cos{(2\pi k'/N)} \langle a_{k'}^\dagger a_{k'}\rangle$ is a contribution that depends on the occupancy level configuration (see App.~\ref{sec:app_B}). With this, we find excellent overlaps in the final ground state with the exact TFD. In \figref{fig:overlap_Hubbard}, we plot the value of the overlap $|\braket{\hbox{GS}(\beta)}{\hbox{TFD}(\beta)}|$  for different values of the coupling strength ratio, $U/t$, for a Hubbard ring of $N=6$ sites. In all the temperature range, the overlap remains very close to 1.

In summary, we have provided a scheme in which, with  controllable  accuracy, the problem of finding the TFD state of a certain Hamiltonian $H$ can be casted to an eigenstate problem in an auxiliary Left-Right coupled system.

\section{Forging a Thermofield Double}\label{sec:forging}

In this section we will address the preparation of the TFD state variationally.
The EF protocol \cite{eddins2022doubling} is specifically  designed to deal with the variational evaluation of the ground state energy in problems which exhibit a natural bi-partition. Define, on the $2N$-qubit  Hilbert space ${\cal H} = {\cal H}_L\otimes {\cal H}_R$, a general variational state written in the Schmidt decomposition form as
\be\ket{\Psi} = \left(U(\btheta)\otimes V(\bvartheta)\right)\sum_{i=1}^{2^N} \lambda_i \ket{b_i}\otimes \ket{ b_i}. \label{TFDpsi}
\ee
Here $\ket{b_i}$ stand for the elements of the computational basis, $\blambda$, $\btheta$ and $\bvartheta$ are {\em sets} of variational parameters and $U$ and $V$ are variational ansatzes. Writing  $\ket{\Psi}$ in this form allows to compute any two-sided cost function $\bra{\Psi}O_L\otimes O_R\ket{\Psi}$ as a  linear combination of  products of one-side expectation values. The number of variational parameters is very large, already $2^N$ just to account for the Schmidt coefficients $\lambda_i$. As mentioned in \cite{eddins2022doubling}, not every problem is well suited to be handled by this strategy. Typically one looks for situations where a truncation to a low number of Schmidt coefficients can be justified.

Indeed, the specific bipartite form of the  TFD  suggests to use a Schmidt like ansatz in order to minimize \eqref{htot}. If after optimization the resulting optimal state $\ket{\Psi_{opt}}\equiv \ket{\Psi(\btheta_{opt}, \bvartheta_{opt})}$ is equal to the  TFD, the Schmidt coefficients must encode precisely the Boltzmann weights $\lambda_i \to \lambda^{opt}_i=e^{-\beta E_i/2}/\sqrt{Z}$. Moreover, $U(\btheta_{opt})$ has to be exactly the matrix that rotates the computational basis into the energy eigenbasis $\ket{E_i}=U(\btheta_{opt})\ket{b_i}$, such that $H \ket{E_i} = E_i\ket{E_i}$.

Motivated by this observation, let us put forward the following TFD-inspired  variational  ansatz
\be
\ket{\Psi(\btheta)} =\sum_{i=1}^{2^N}  \frac{e^{-\beta\tilde E_i(\btheta)/2}}{\sqrt{Z(\beta,\btheta)}}  \ket{f_i(\btheta)}\otimes \ket{ f^*_i(\btheta)}, \label{wave_fun}
\ee
where $\ket{f_i(\btheta)}=U(\btheta)\ket{b_i},\ket{f^*_i(\btheta)}=U^*(\btheta)\ket{b_i}$,\footnote{Notice that this complex conjugation is trivial to implement provided that the structure of the circuit is known. In particular, if $U$ is only composed by Pauli rotation gates, $U^*(\theta)=U(-\theta)$.}  the energy estimators $\tilde{E}_i(\btheta)=\bra{f_i(\btheta)} H \ket{f_i(\btheta)}$, and the normalization factor $Z(\beta,\btheta) = \sum_k e^{-\beta \tilde E_k(\btheta)}$.
Notice the exponential reduction in the number of parameters that occurs upon replacing the Schmidt coefficients $\lambda_i$  in favor of the estimators $\tilde{\lambda}_i = e^{-\beta \tilde E_i(\btheta)}/\sqrt{Z(\btheta)}$.

\begin{figure}[t]
\centering
\includegraphics[scale=0.36]{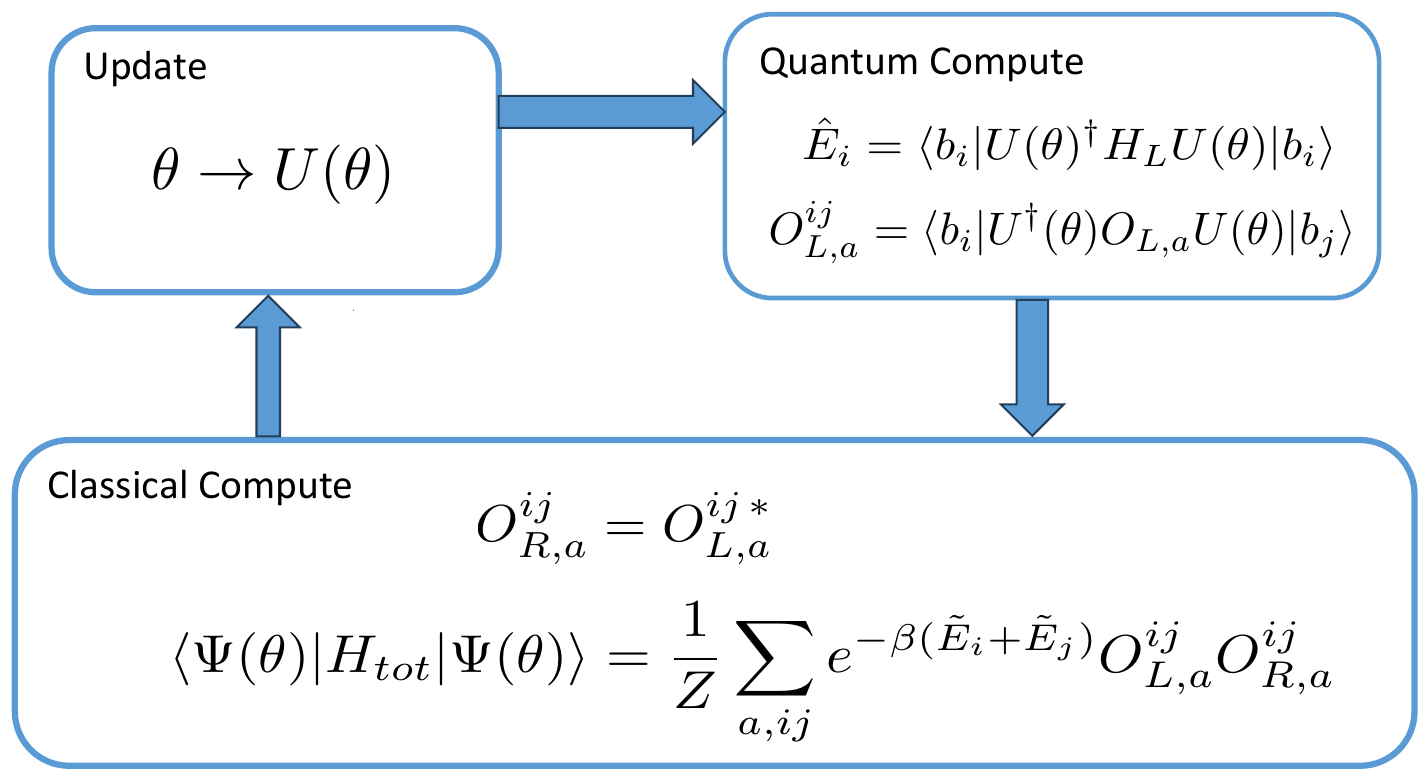}
\caption{For a Hamiltonian of the form 
$H_{tot} = H_L + H_R + H_{LR}= \sum_a O_{L,a}\otimes O_{R,a}$ only one sided expectation values need to be evaluated quantumly. The classical part composes and minimizes  the global energy cost function $\langle H_{tot}\rangle$   }
\label{fig:workflow}
\end{figure}

\begin{figure*}[t]
\begin{center}
\includegraphics[scale=0.45]{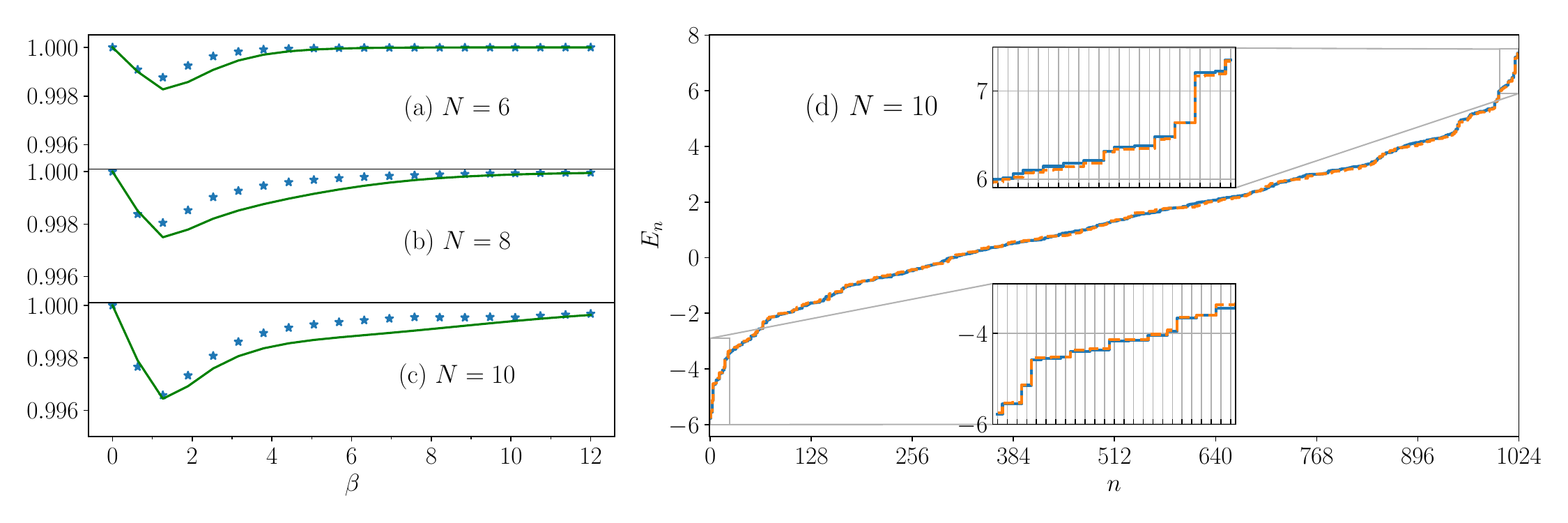}~~~~
\caption{Simulation results for the Hubbard model in frequency space with $t=1$, $\epsilon=0$, $U=0.5$. (a), (b), (c) Comparison between overlaps, $\lvert \braket{\Psi_{opt}}{\hbox{TFD}}\rvert$ (blue stars) and $\lvert \braket{\hbox{GS}}{\hbox{TFD}}\rvert$ (green line) at $N=6,8,10$.  (d) Spectrum obtained via exact diagonalization (blue lines) and variationally (orange dashed lines) at $\beta=1.26$, where the worst value of the overlap is obtained. Insets show the lower and upper parts of the spectrum with more detail. Overall, the variational energies show strong agreement with the exact ones. }
\label{fig:simulations}
\end{center}
\end{figure*}
The evaluation of the  expectation value of any operator  $O_L\otimes O_R$ becomes a weighted sum involving  products of $N$-qubit matrix elements
\begin{equation}
\begin{split}
\bra{\Psi(\btheta)}O_L\otimes O_R&\ket{\Psi(\btheta)} =  \sum_{i,j=1}^{2^N} \frac{e^{-\beta(\tilde E_i(\btheta)+\tilde E_j(\btheta))/2}}{Z(\beta,\btheta)}  \\
&\times 
\bra{f_i(\btheta)}O_L \ket{f_j(\btheta)}\bra{f^*_i(\btheta)} O_R \ket{f^*_j(\btheta)}. 
\end{split} \label{expectationEF}
\end{equation}
Given that the  cost function  is  $\bra{\Psi(\btheta)}H_{tot}(\beta)\ket{\Psi(\btheta)}$ we can write $H_{tot} = \sum_a O_{L,a} \otimes O_{R,a}$ and it is clear that, at each round, everything is set in place and the workflow is consistent (see  \figref{fig:workflow}). 
Remember that $\beta$ is not a variational parameter, as it is fixed by $H_{tot}(\beta)$. 

As it is well known in the context of the circuit cutting technology, the price for halving the circuit width is paid in the exponentially larger amount of expectation values one has to compute. The usual procedure (see \cite{eddins2022doubling}) to alleviate the protocol is to invoke a good reason to truncate the  Schmidt coefficients to a handful of relevant ones. In the present case, the exponentially decaying nature of the Boltzmann factors, $\lambda_i \propto e^{-\beta E_i/2}$, easily justifies to truncate the sum to the lower part of the spectrum. Remarkably, this  makes the procedure work much better in the low-temperature regime $\beta \gg 1$. Precisely the opposite regime  usually addressed by other methods, like QAOA \cite{wu2019variational} or QITE \cite{Motta_2019,McArdle_2019}. In App.~\ref{sec:app_C} we analyze how the truncation impacts the optimization and how $\beta$ can be used as a hyperparameter that controls the complexity of the problem.

It is natural to expect that the optimization will attract the variational state $\ket{\Psi(\btheta)}$ towards the  TFD-like state closest to $\ket{\hbox{GS}}$. Provided that $|\braket{\hbox{GS}(\beta)}{\hbox{TFD}(\beta)}|\approx1$, then also $|\braket{\Psi_{opt}(\beta)}{\hbox{TFD}(\beta)}|\approx1$. In Sec. \ref{sec:results}, we will provide numerical evidence supporting this expectation. 

Once the minimization is completed, the outcome yields the full spectrum of energies $E_i$ as well as the unitary   that diagonalizes $H$, i.e. $\ket{E_i} = U(\btheta_{opt})\ket{b_i}$. These  ingredients are sufficient to compute thermal expectations values. 

Still, one may be  interested in a circuit whose output is the TFD state itself. This is  the case, for example, in wormhole inspired teleportation protocols \cite{maldacena2018eternal, Su_2021}. For that, we must supply a $2N$ circuit that finishes the job. The only remaining part is a $N$-qubit circuit that loads the previously obtained Boltzmann coefficients in the computational basis
\begin{equation}
U_{\lambda}\ket{0}=\sum_i \lambda_i \ket{b_i} \approx \sum_i \tilde{\lambda}_i \ket{b_i} \, .\label{loading}
\end{equation}
This state can then be duplicated and plugged into \eqref{TFDpsi} to find the desired TFD state.
While the probability loading problem is well studied in the literature following standard techniques \cite{grover2002creating, marin2023quantum}, we will provide a solution which, again, uses the exact free fermion case as an initial approximation. 

Consider, for definiteness, the following $N=3$ circuit 
\begin{center}
\begin{quantikz}[row sep=0.2cm]
&\gate{R_Y(2\varphi_0)}&\gate[3]{\tilde{U}_\lambda(\bphi)}& \qw\\
&\gate{R_Y(2\varphi_1)}& &\qw\\
&\gate{R_Y(2\varphi_2)}& &\qw\\
\end{quantikz}
\end{center}

The concrete values of the rotation angles are related to the weights $\omega_i$ of the free fermion Hamiltonian $H_2$ in \eqref{hamil} as follows
$$
 \varphi_i(\omega_i) =\arctan (e^{-\beta \omega_i/2}).
$$
In the free case, $H = H_2$, the first layer of $R_Y$ rotations suffices to perform the loading task \eqref{loading} exactly as a Bogoliubov transformation (see App.~\ref{sec:app_A}). For the interacting case, $H = H_2 + H_4$,   the circuit needs to be supplemented with an additional variational subcircuit $\tilde U_\lambda(\bphi)$, for which the previous layer provides a {\em warm start} at $\tilde U_\lambda(0)=I$.\footnote{Of course, in the interacting case, the free piece should be optimized first, $\omega_i \to \tilde \omega_i =\omega_i + \delta_i$ by, for example, Mean Field methods, to maximize the effectiveness of the $H_{LR}$.}
In order to optimize $\tilde U_\lambda(\bphi)$ we  propose the cost function 
\begin{equation}
    \mathcal{C}(\bphi) = \sum_{i=1}^{2^N} \left|\lambda_i^2 - p_i(\bphi)\right|\, ,
\end{equation}
where $p_i(\bphi)$ are the probabilities of measuring each of the states of the computational basis of the circuit, and $\lambda_i\approx \tilde{\lambda}_i$ are the Schmidt coefficients obtained in the previous step. Same as before, the terms in this sum decrease as $\sim e^{-\beta E_i/2}$ and should be truncated from below to a finite number, in order to deal with a feasible optimization problem.
In App.~\ref{sec:app_D}, we do a more in depth analysis of this proposal and show that the optimization yields overlaps above $0.999$ for the Hubbard model.

When all the parameters have been optimized, we can finally write down the generating circuit for the TFD state. 

\begin{center}
\begin{quantikz}[row sep=0.15cm,column sep=0.2cm]
&\gate{R_Y(2\varphi_0(\tilde \omega_0))}&\gate[3]{\tilde{U}_\lambda(\bphi_{opt})}& \ctrl{3} & \qw & \qw &\gate[3]{U(\btheta_{opt})^{\ }} & \qw\\
&\gate{R_Y(2\varphi_1(\tilde \omega_1))}& \qw &\qw&\ctrl{3} & \qw& &\qw\\
&\gate{R_Y(2\varphi_2(\tilde \omega_2))}& \qw &\qw& \qw&\ctrl{3} & &\qw\\
&\qw&\qw&\targ{}&\qw&\qw& \gate[3]{U(\btheta_{opt})^*}& \qw  \\
&\qw&\qw&\qw&\targ{}&\qw& & \qw\\
&\qw&\qw&\qw&\qw&\targ{}&&\qw
\end{quantikz}
\end{center}

\section{Simulations and Results} \label{sec:results}

Selecting an adequate variational ansatz for the circuit $U(\btheta)$ is necessary in order to forge the TFD as the ground state of $H_{tot}$ \cite{eddins2022doubling}.  As any other variational algorithm,  it can suffer from trainability issues such as local minima or barren plateaus \cite{larocca2024review}. Moreover,  it also has to be expressive enough for the ansatz to approximate the unitary change  to the  basis $\ket{E_i}$ of eigenstates of $H$.

With these caveats in mind, we opt for a Hamiltonian Variational Ansatz (HVA) \cite{Wecker2015, Wiersema2020HVA} from which we can generate the vectors  $\ket{f_i({\theta})}=U(\btheta)\ket{b_i}=\prod_l^\mathcal{L} \left( \prod_s e^{-i\theta_{s,l} h_s} \right) \ket{b_i}$, directly from the elements of the computational basis, which are the eigenstates of $H_2$. The product is over sets $h_s$ made of commuting operators such that  $H=\sum_s h_s$, while  $[h_s,h_s']\neq 0$. By construction, in our case $h_{0}$ is the Jordan-Wigner transform of $H_2$. The index $l$ iterates over the number of layers $\mathcal{L}$.

In \figref{fig:simulations}, results are given for the optimizations using scipy's BFGS optimizer and $\mathcal{L}=1$. The Hamiltonian is the one given in \eqref{Hub_k} with parameters $t=1$, $\epsilon_0=0$ and $U=0.5$. Overlaps and the resulting spectrum have been used as figures of merit. For the overlaps, we find that generally $\lvert \braket{\Psi_{opt}}{\hbox{TFD}} \rvert \geq \lvert \braket{\hbox{GS}}{\hbox{TFD}} \rvert$, which can be interpreted as $\ket{\Psi_{opt}}$ being even closer  to $\ket{\hbox{TFD}}$ than $\ket{\hbox{GS}}$ itself. Additionally, for $N=10$, we compare the spectrum $E_i$ obtained from exact diagonalization, and the one obtained variationally $\tilde{E}_{i}(\btheta_{opt})$ at $\beta=1.26$, which is the worst value for the overlap. We see that they match almost perfectly in both the low tail and top head of the spectrum. We do not find any trainability issues in these optimizations as long as the parameters are initialized precisely at zero value. In this regime, the starting point is the TFD state of $H_2$ which due to the perturbative nature of our protocol is a \textit{warm start} for the optimization. Random initialization, on the contrary, does lead to optimization problems that prevent convergence. This is not surprising, as even though the ansatz is constrained to a $N$ qubit subspace, the optimization occurs in the complete $2N$ space.\footnote{The relevance of a  heuristic preparation is a subject of intense debate \cite{Lee2023,Grimsley2023,Cerezo2023}.}

\section{Conclusions and comparison with other proposals}\label{sec:conclusions}

We have proposed a variational construction of the TFD that approximates this state as the ground state of a suitably engineered Hamiltonian. 
The range of applicability  restricts to systems that can be mapped onto fermionic Hamiltonians, and builds perturbatively upon the fact that the free fermion case is exactly solvable (it would be interesting to leverage this statement with some interaction distance \cite{Pachos_2018}). On general grounds we expect the method to provide good fidelities up to a coupling strength where a ground state level crossing or, in more generality, a quantum phase transition may occur. 

Our proposal involves two circuits in sequence, both of width $N$, the dimension of the Hilbert space. In the first one, the full spectrum and the rotation matrix that diagonalizes the Hamiltonian are obtained. The variational ansatz adapts naturally to the use of EF. The exponentially decaying structure of the Schmidt coefficients, $\lambda_i =e^{-\beta E_i/2}/\sqrt{Z}$, allows for a truncation to the lowest part of the spectrum by neglecting them below some cutoff. Depending on the case, one may want to run or not the second circuit that actually builds the TFD proxy. If not required, this implies a substantial saving in computational cost.

The fact that our algorithm works better for low temperatures is remarkable, and worth considering when comparing it with other proposals which, generically start from the simple, infinite temperature, maximally mixed. For example in \cite{wu2019variational}, a variational construction of the thermofield double is proposed starting from that state and performing a QAOA evolution.\footnote{The failure of the method when trying to reach low temperatures is scrutinized in figure 3c in that paper.} Also in the QITE proposal \cite{Motta_2019,McArdle_2019}, the euclidean evolution time is longer the lower the target temperature and, hence, harder to simulate. 

The ground state preparation problem could also be tackled using an adiabatic approach \cite{Farhi2000}. A posible method involves adiabatically evolving the free TFD state $\ket{0(\beta)}$ (see App.~\ref{sec:app_A} and ~\ref{sec:app_C}) under a time-dependent Hamiltonian $H(t)=\left(1-t/T\right)H_{tot}\left(\beta, U=0\right)+t/TH_{tot}\left(\beta, U\right)$. This approach requires the total evolution time $T$ to be sufficiently large so that the process remains slow, and the gap of $H(t)$ does not vanish. Under these conditions, the adiabatic protocol should work as intended. However, implementing such time evolution on a digital quantum computer presents significant challenges.
To execute this process digitally, $T$ must be discretized into $N$ time steps of size $\Delta t = T/N$. The evolution operator can be approximated \cite{Poulin2011} as $U(t)=\mathcal{T}e^{-i\int H\left(t^\prime\right) dt^\prime}\approx \prod_k^N e^{-iH\left(t_k\right)\Delta t}\approx \prod_k^N \left(\prod_s e^{-iH_s(t_k)\Delta t/M}\right)^M$. Here, the Hamiltonian at each time step is expressed as $H(t_k)=\sum_sH_s(t_k)$, and the Lie-Trotter formula is used with $M$ iterations to approximate its evolution. For this approximation to be valid, the condition $\Delta t/M \lll 1$ must hold, which implies that $NM$ must be large. Compared to the HVA, the resulting circuit for the adiabatic protocol is approximately $NM$ times deeper. This makes it impractical for non-fault-tolerant quantum computers, where running deep circuits is infeasible. Consequently, variational methods are often preferred, despite their associated optimization challenges. In our specific setup, however, optimization issues do not seem to arise. Furthermore, our variational protocol has an additional advantage: it preserves the circuit structure of the TFD. This is critical for obtaining both the energy eigenvalues and the diagonalizing operator—key features that would be lost in adiabatic processes.

Finally the fact that, in our case, the quadratic part is exactly solvable is extremely important. It provides a physically motivated initial state ({\em warm start}). This implies that  in both variational circuits,  the initial value of the parameters is not random, but precisely $\btheta = \bphi=0$. The overlap of this initial state with the final solution is significant even for rather high values of the interaction strength. We have seen that this  fact, together with the effective implementation of a Hamiltonian Variational ansatz, makes  the convergence of the optimization process seamless. Presumably, it will also affect the scaling in circuit complexity, something to study in the near future.

While completing this work we came across \cite{consiglio2023variational} that contains a proposal which bears strong resemblance with ours.  The variational proposal  is very compact as it involves optimizing altogether a $2N$-qubit circuit, with  parameters $(\btheta, \bphi)$ playing essentially the same roles as the ones in our approach.  The full circuit is coupled, so all the parameters need to be optimized together. 
Upon convergence, the outcome is a circuit that composes the correct TFD state.
As usual in  variational circuits, the initial state has an overlap with the searched ground state that vanishes exponentially fast with the size of the system, and this is behind the exponential scaling in complexity that affects them generically. 
In contrast, in our proposal,  the tasks of obtaining the eigenspectrum and that of constructing the TFD have been decoupled from one another.  Moreover, and thanks to the entanglement forging trick, this has the benefit of dealing always  with width $N$, instead of $2N$,  optimization circuits.

\section{Acknowledgments}
We would like to thank Diego Porras, Sebastián V. Romero and Luca Tagliacozzo for interesting discussions. 

The work of J.SS. and J.M.  was supported by Xunta de Galicia (Centro Singular de Investigacion de Galicia
accreditation 2019-2022) and by the Spanish Research State Agency under grant PID2020-114157GB-I00, and by the European Union FEDER. The work of J.SS. was supported by  MICIN through the European
Union NextGenerationEU recovery plan (PRTR-C17.I1), and by the Galician Regional Government through the
“Planes Complementarios de I+D+I con las Comunidades Autónomas” in Quantum Communication. D.F. was supported by Axencia Galega de Innovación through the Grant Agreement ``Despregamento dunha infraestructura baseada en tecnoloxías cuánticas da información que permita impulsar a I+D+I en Galicia'' within the program FEDER Galicia 2014-2020. Simulations in this work were performed using the Finisterrae III Supercomputer, funded by the project CESGA-01
FINISTERRAE III. D.H.M. acknowledges financial support from CEA's Science Impulse Program and the France 2030 HQI project (ref. ANR-22-PNCQ-0002).

\appendix
\renewcommand{\thefigure}{\,A\arabic{figure}}
\setcounter{figure}{0}   

\section{Quadratic Hamiltonian}\label{sec:app_A}

Any free fermion Hamiltonian is quadratic in the fermion operators, and can be brought to a diagonal form
$$
H =  \sum_{k=1}^n \omega_k  a^\dagger_k a_k,
$$
where $\omega_k$ are  real quantities which we will assume positive without loss of generality.  In this case the ground state $\ket{0}$ of $H$ is given by the Fock vacuum $a_k\ket{0} = 0$.

 Following  \cite{Takahashi}, in order to build the TFD state we double the Hilbert space $a_k \to a_k^L,a_k^R$, and define
$$
G_F(\beta)  = i\sum_{k=1}^n \varphi_k(\beta,\omega_k) ( a^L_k a^R_k + a^{L\dagger}_k a^{R\dagger}_k ) 
$$
where $\varphi_k = \arctan e^{-\beta \omega_k/2}$. Now we can define the rotated ground state 
\be
\ket{0(\beta)}~ = e^{-i  G_F(\beta )}\ket{0} =  \prod_{k=1}^n (u_k+ v_k a^{L\dagger}_k a^{R\dagger}_k) \ket{0}\, ,\label{zerobeta1}
\ee
where $u_k = \cos\varphi_k$ and $v_k = \sin\varphi_k$.
An explicit expansion of \eqref{zerobeta1} shows that, indeed, $\ket{0(\beta)} $ is a thermofield double at $T = \beta^{-1}$
\be
\begin{split}
\ket{0(\beta)} = \frac{1}{\sqrt{Z}} & \left( 1+  \sum_k e^{-\beta \omega_k/2} a_k^{L\dagger} a_k^{R\dagger} \right. \\
& \left. -   \sum_{k\neq l} e^{-\beta (\omega_k+\omega_l)/2} a_k^{L\dagger} a_l^{L\dagger} a_k^{R\dagger} a_l^{R\dagger} + ... \right) \ket{0}
\end{split}
\label{zerobeta2}
\ee
Consider now the Bogoliubov-transformed Fock operators
\beqa
\tilde a^L_k(\beta) &=&  e^{-iG_F} a^L_k e^{iG_F} = u_k(\beta)  a^{L}_k - v_k(\beta) a^{R\dagger}_k   \nonumber \\
\tilde a^R_k(\beta) &=&  e^{-iG_F} a^R_k e^{i G_F} = u_k(\beta) a^R_k  + v_k(\beta) a^{L\dagger}_L  \, .\nonumber
\eeqa
With them, we can define the $(2n+1)$-parameter family of deformed Hamiltonians
\be
\nonumber
\begin{split}
\tilde H(\beta, L, R) & = e^{-iG_F(\beta)} H e^{iG_F(\beta)} \\
& =  \sum_i \left( L_i\,  \tilde a_i^{L\dagger}(\beta) \tilde a_i^L(\beta)  + R_i\,   \tilde a_i^{R\dagger}(\beta)  \tilde a^R_i(\beta) \right) \, .
\end{split}
\ee
From the construction it follows that, {\em for any choice of} $L_i, R_i$, the rotated Fock vacuum $\ket{0(\beta)}$ in \eqref{zerobeta2}  is the ground state of the rotated Hamiltonian, $\tilde H(\beta, L,R)$. Inserting the explicit form of the transformed oscillators we get
\be
\begin{split}
\tilde H(\beta;L,R) =   \sum_{i} & \left[  \rule{0mm}{5mm}(L_i\, u_i^2 - R_i v_i^2) \, a_i^{L\dagger} a_i^L \right.  \\
& \left. + (- L_i\, v^2_i + R_i u^2_i)\, a_i^{R\dagger}a_i^R   \right. \\ 
\rule{0mm}{5mm}
&\left.   + (L_i\,   + R_i ) v_i u_i \, (   a_i^L a_i^R +  \,  a_i^{R\dagger}   a_i^{L\dagger} ) \right. \\  & \left. +( L_i + R_i )v^2_i \right] \, .
\end{split}
\label{cont_fam}
\ee
Cases of particular interest are the following ones
\begin{itemize}
\item $L_i= -R_i=\omega_i$. Any reference to $\beta$  disappears and we obtain
\beqa
\tilde H&=& \sum_{i=1}^n \omega_i (a_i^{L\dagger} a_i^L -  a_i^{R\dagger} a_i^R )  = H_L - H_R 	\label{lminusr}
\eeqa
\item $L_i= R_i= \displaystyle \frac{\omega_i}{(u_i)^2 -   (v_i)^2} $
\be
\begin{split}
 \qquad \tilde H(\beta)= \sum_i& \left[  \omega_i (a_i^{L\dagger} a_i^L +  a_i^{R\dagger} a_i^R) \right. \\
& \left. + 2 \mu_i(\beta)  \, ( a_i^L a_i^R  +    a_i^{R\dagger} a_i^{L\dagger}) + 2 C_i(\beta) \right] 
\end{split}
\label{lplusr}
\ee
where the dependence upon $\beta$ appears only in the mixed term coupling and zero point energy
$$
\mu_i(\beta)  =  \frac{ v_i u_i \omega_i}{u_i^2 -   v_i^2}~~~~,~~~~C_i(\beta) = \frac{\omega_i v_i^2 }{u_i^2 -   v_i^2}
$$

$\bullet$ A third possibility makes connection with the entanglement cloning Hamiltonian \cite{Hsieh_2016}.
Take for this
$$
L_i = \frac{u_i^2}{u_i^2- v_i^2}\omega_i ~~~~,~~~~ R_i = \frac{v_i^2}{u_i^2- v_i^2}  \omega_i
$$
and a straightforward computation gives
\be
\begin{split}
\quad \, \, \,  \tilde H(\beta)  = \sum_i & \Bigr[ \omega_i a_i^{L\dagger}a_i^L
+ \mu_i(\beta) (a_i^L a_i^R  + a_i^{R\dagger} a_i^{L\dagger}) \\
 & + \, C_i(\beta) \, \Bigr]
\end{split}
\label{lonly}
\ee
\end{itemize}
This proves that, indeed,  the TFD is the exact ground state of a continuous family of Hamiltonians \eqref{cont_fam}, of which we show three particular cases,  \eqref{lminusr} \eqref{lplusr} and \eqref{lonly}, which are very simply related to the original Hamiltonian.

Notice that there are some minus signs that appear in the TFD in \eqref{zerobeta2}. Their origin can be traced back  to the fermionic statistics. They can be reabsorbed in a redefinition of the $R$ basis. A practical way to obtain all the signs posive is  skipping the $Z$-strings when applying the Jordan Wigner transform to the $H_{LR}$ Hamiltonian.

\section{Mean-Field Approximation}\label{sec:app_B}
\renewcommand{\thefigure}{\,B\arabic{figure}}
\setcounter{figure}{0}   
As we have shown, in the non-interacting case
\be
    H_{L}=\sum_i \omega_i a_i^\dagger a_i, 
\ee
the Thermofield Double  is the ground state  of $H_{tot}(\beta)=H_L+H_R+H_{LR} (\beta)$  when the coupling Hamiltonian between $H_L$ and $H_R$ is  given by
\be
    H_{LR} (\beta)=\sum_i \mu_i (\beta )(a_i^{L} a_i^{R}-a_i^{\dagger L} a_i^{\dagger R}),
\ee
and the weights $\mu_i(\beta)=\frac{\omega_i}{2 \sinh{\beta \omega_i /2}}$. However, relevant problems in physics usually involve complex Hamiltonians with quartic and higher-order terms
\begin{equation}
    H_L=\sum_i \omega_i a^\dagger_i a_i + \sum_{ijkl} U_{ijkl} a^\dagger_i a^\dagger_j a_k a_l + ...
\label{H_generic}
\end{equation}
for which our prescription can  provide accurate approximations.

\subsection{Spinless 1D Hubbard model}

\begin{figure*}[t]
\includegraphics[scale=0.45]{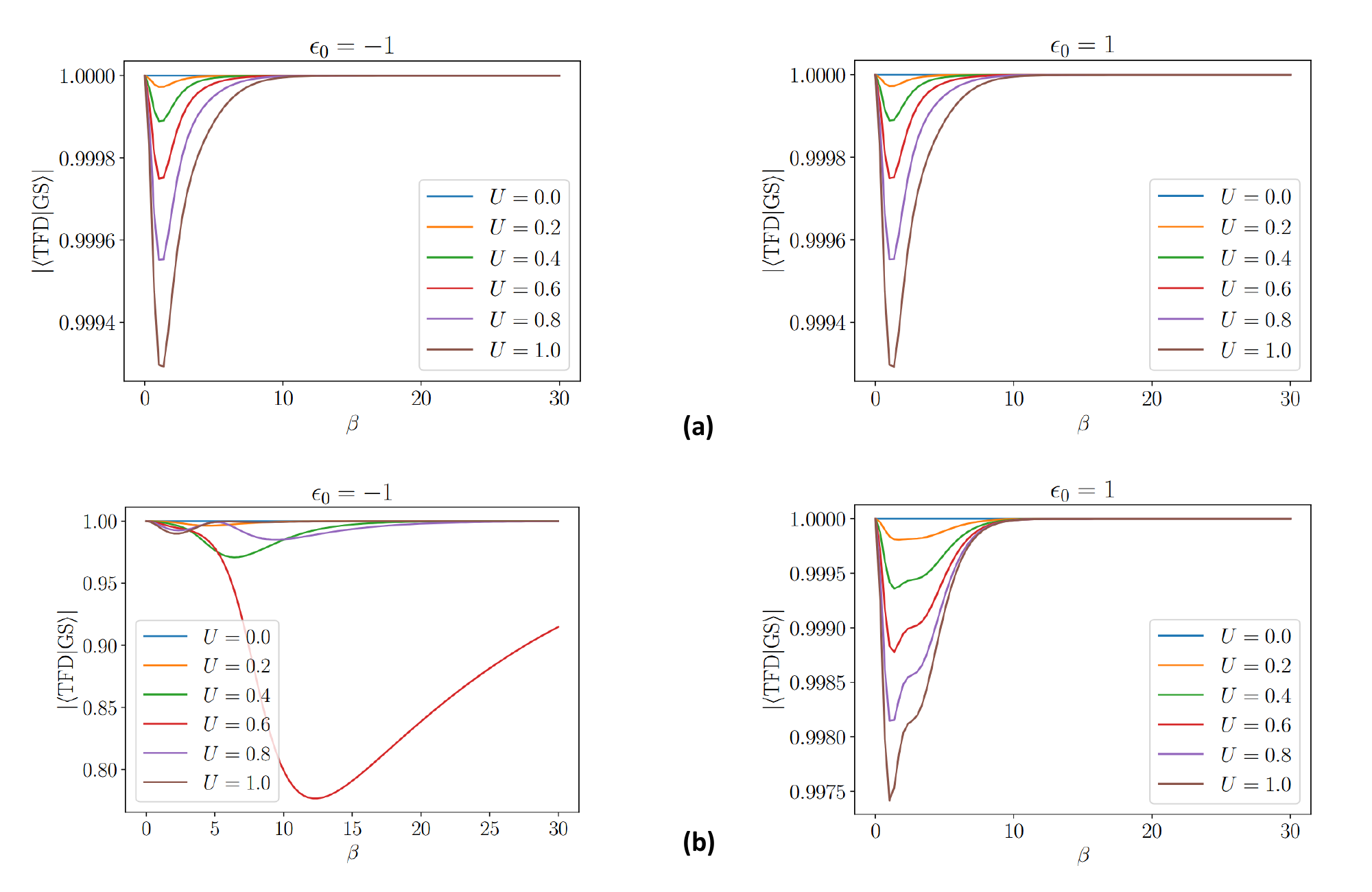}
\caption{Overlap between the TFD and the GS for the 1D spinless Hubbard model for $\epsilon_0=\lbrace-1,1 \rbrace$ considering (a) only off-diagonal contributions in the second term of (6), and (b) the full Hamiltonian. The latter case allows interactions to modify the diagonal part, making the non-interacting energies $\omega_k$ inefficient in some cases to achieve a good estimation of the weights in $H_{LR}$. In contrast, in (a), the shortened Hubbard term does not modify the $\omega_k$, and we can control the deviation from the exact case ($U=0$) through the ratio between $U$ and the hopping constant $t$. Plots show the overlaps for two coupled $N=4$ chains using $t=1$ and varying $U$. }
\label{B1}
\end{figure*}

To demonstrate this, we consider a 1D spinless version of the Hubbard model
\begin{equation}
    H_L=\sum_i \epsilon_0 a^\dagger_i a_i -t \sum_i (a^\dagger_i a_{i+1}+a^\dagger_i a_{i-1}) + U \sum_i a^\dagger_i a_{i} a^\dagger_{i+1} a_{i+1}
\label{Hubb_real}
\end{equation}
that can be written in the same form as \eqref{H_generic} by assuming periodic boundary conditions and performing a Fourier transformation. This is done by substituting the creation/annihilation operators
\begin{equation}
    a_i^\dagger=\frac{1}{\sqrt{N}} \sum_k e^{-ikr_i} a^\dagger_k, \qquad \qquad  a_i=\frac{1}{\sqrt{N}} \sum_k e^{ikr_i} a_k \, ,
\end{equation}
which leaves the previous Hamiltonian \eqref{Hubb_real} as
\begin{equation}
    H_L=\sum_k \omega_k a_k^\dagger a_k +\frac{U}{N}\sum_{k,p,q} e^{-i\frac{2\pi}{N}q}a_{k+q}^\dagger a_k a^\dagger_{p-q}a_p \, ,
\label{Hubb_k}
\end{equation}
where $\omega_k=\epsilon_0 -2t \cos{\left(\frac{2\pi}{N} k\right)}$ with $k=0,1,2..,N-1$. As we proved, for $U=0$, the overlap between the TFD and the ground state of $H_{tot}$ is one for every $\beta$. For $U\neq 0$, two situations can take place. First, the interaction does not modify the non-interacting energies $\omega_k$, i.e., its corresponding matrix is completely off-diagonal (\figref{B1}a). Then, the overlap between the TFD and the GS will depart from 1 in a controllable way as we increase  $U$. Second, interactions modify the diagonal part, and hence, the energies $\omega_k$ could not be  good candidates to estimate $H_{LR}$. That is the case of \eqref{Hubb_k} for $q=0$ and $k+q=p$. In this situation, overlaps could be worse even in a weak coupling regime, as we see in \figref{B1}b. To address this, it is necessary to estimate the contribution of $U$ in the non-interacting energies $\omega_k$.

The most natural way is to perform a mean-field approximation in the Hubbard Hamiltonian. That is, treating correlations with the other particles as a mean density and transforming the Hamiltonian into a quadratic one with average densities. The essence of this approximation relies on the fact that density operators $a_{k}^\dagger a_{k'}$ deviate in a small quantity from their average values $\langle a_k^\dagger a_{k'} \rangle$. Therefore, it is adequate in our case, where we are not interested in strong perturbations to the free fermion regime. It is important to note that we are not using the mean-field approximation to simplify our problem but to estimate correctly the energies, $\omega_k(\epsilon_0,t) \rightarrow \tilde\omega_k(\epsilon_0,t,U)=\omega_k +\delta\omega_k$, from which we generate $H_{LR}$.

\begin{figure}
\includegraphics[scale=0.45]{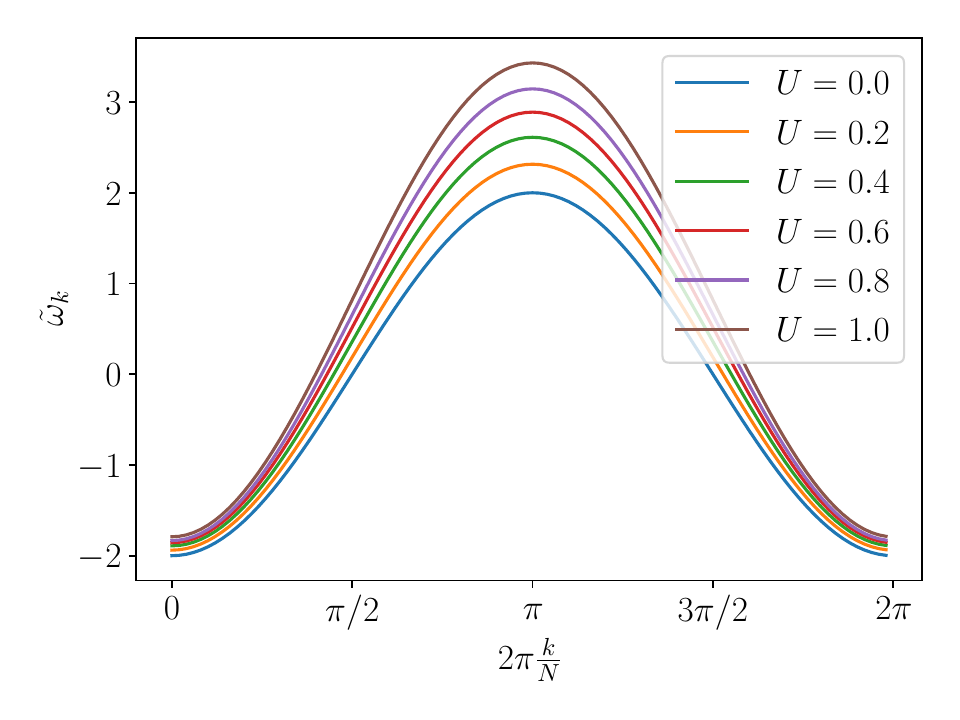}
\caption{Mean-field energies $\tilde\omega_k$ \eqref{Mean-field energies} for different values of $U$ and $N=100$.}
\label{B2}
\end{figure}

\begin{figure*}
\centering
\includegraphics[scale=0.45]{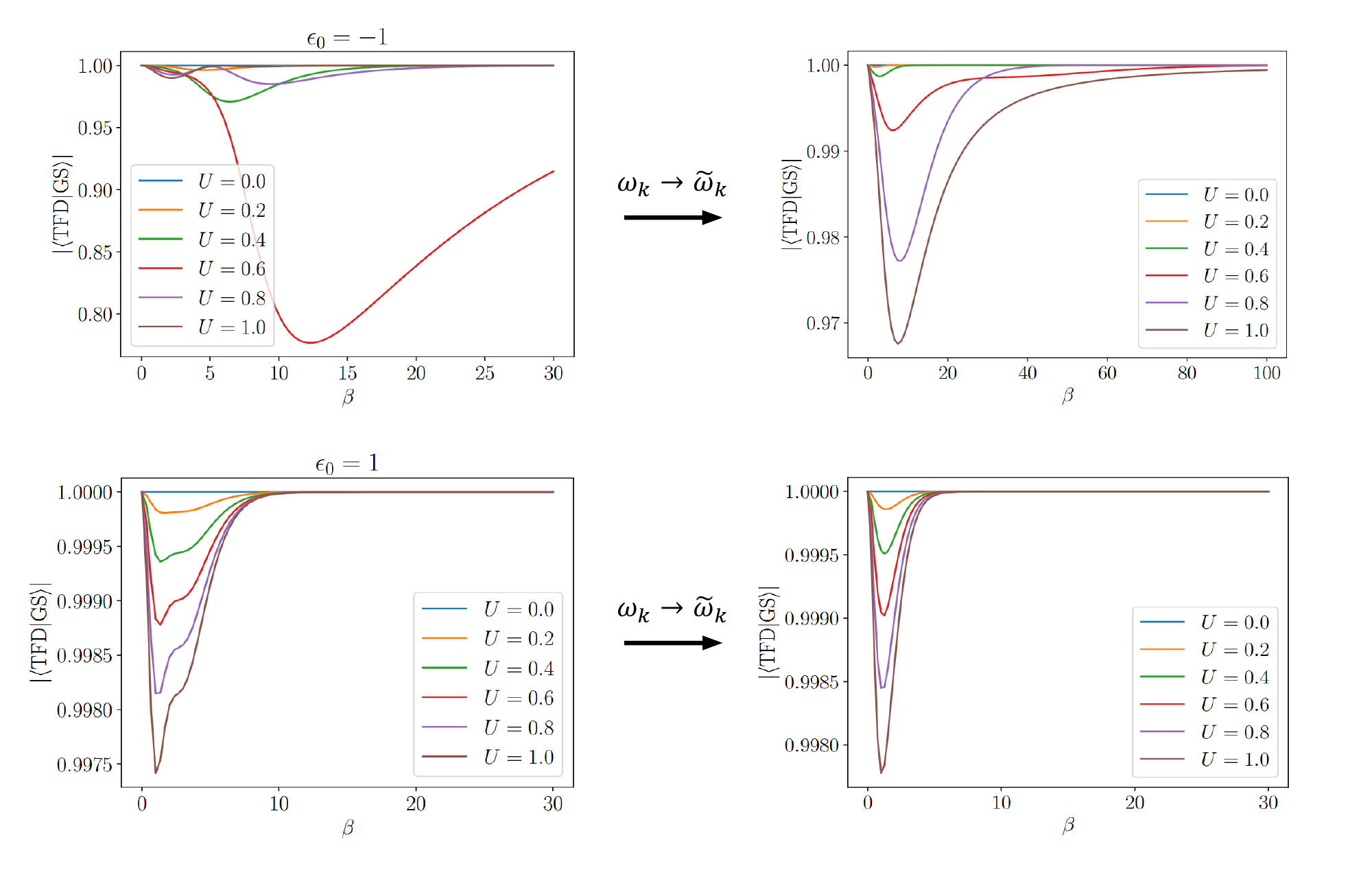}
\caption{ Overlap for the 1D spinless Hubbard model ($\epsilon_0=\lbrace-1,1 \rbrace$, $t=1$, $2N=8$). Left: Using the non-interacting energies $\omega_k=\epsilon_0-2t\cos{\left(\frac{2\pi}{N}k\right)}$. Right: Using their corrected values $\tilde\omega_k$ within the mean-field approximation.}
\label{B3}
\end{figure*}

The mean-field approximation of \eqref{Hubb_k} gives the following Hamiltonian \cite{bruus2004many}
\begin{equation}
\begin{split}
    H^{MF}=\sum_k \omega_k a_{k}^\dagger a_k & + \frac{U}{N} \sum_{k,p,q} e^{-i\frac{2\pi}{N}q}  a_{k+q}^\dagger a_{k} \langle a_{p-q}^\dagger a_p\rangle \\
    &+ \frac{U}{N} \sum_{k,p,q} e^{-i\frac{2\pi}{N}q}  a_{p-q}^\dagger a_{p} \langle a_{k+q}^\dagger a_k\rangle  \\
    & - \frac{U}{N} \sum_{k,p,q} e^{-i\frac{2\pi}{N}q}   \langle a_{k+q}^\dagger a_p\rangle a_{p-q}^\dagger a_{k} \\
    &- \frac{U}{N} \sum_{k,p,q} e^{-i\frac{2\pi}{N}q}   \langle a_{p-q}^\dagger a_k\rangle a_{k+q}^\dagger a_{p}\\
    & - \frac{U}{N} \sum_{k,p,q} e^{-i\frac{2\pi}{N}q}   \langle a_{k+q}^\dagger a_k\rangle \langle a_{p-q}^\dagger a_p\rangle \\
    &+\frac{U}{N} \sum_{k,p,q} e^{-i\frac{2\pi}{N}q}   \langle a_{k+q}^\dagger a_p\rangle \langle a_{p-q}^\dagger a_k\rangle .\label{mean_field1}
\end{split}
\end{equation}
Assuming  translational invariance, the  average densities can be written as  $\langle c^\dagger_k c_k' \rangle= \delta_{k.k'} \langle n_k\rangle$ \cite{bruus2004many}. The odd terms in $q$ cancel in the sums, and the mean-field Hamiltonian turns out to be
\begin{equation}
\begin{split}
    H^{MF}&=\sum_k \omega_k a_{k}^\dagger a_k + \frac{2U}{N} \sum_{k}  a_{k}^\dagger a_{k} \sum_p \langle n_p \rangle \\ 
    &- \frac{2U}{N} \sum_{k} \cos{\left(\frac{2\pi}{N}k\right)}  a_{k}^\dagger a_{k} \sum_p \cos{\left(\frac{2\pi}{N}p\right)}\langle n_p \rangle \\
    & - \frac{U}{N} \sum_{k} \langle n_k \rangle \sum_p \langle n_p \rangle \\
    &+\frac{U}{N} \left(\sum_{k} \cos{\left(\frac{2\pi}{N}k\right)} \langle n_k \rangle \right)^2\, , 
\end{split}\label{mean_field2}
\end{equation}
from where the modified energies can be read off 
\begin{equation}
\tilde\omega_{k}=\omega_k +2U\rho -\frac{2U}{N}  \cos{\left(\frac{2\pi}{N}k\right)} \sum_p \cos{\left(\frac{2\pi}{N}p\right)}\langle n_p \rangle 
\label{Mean-field energies}
\end{equation}
being $\rho=\frac{N_e}{N}=\frac{1}{N} \sum_p \langle n_p \rangle$ the density of particles.  Two equivalent routes can be applied to solve this problem i) determining energies and densities self-consistently $\sum_k \langle n_k \rangle= \sum_k \theta(\mu-\tilde\omega_k)$  or ii) minimizing the free energy with respect to the average densities \cite{bruus2004many}. In our case, we choose to minimize the energy numerically by finding the optimal configuration  \cite{Pavarini:819465}. That is, we run over $N_e$, filling the $N_e$-lowest energy levels, and we identify the lowest energy configuration. Once the modified energies $\tilde\omega_k$ (\figref{B2}) have been obtained, we recompute the overlaps. In general, they show a substantial improvement when compared to the case of using the non-interacting energies $\omega_k$. \figref{B3}  shows the numerical results for the previous cases. Finally, note that the overlaps are not invariant under $\epsilon_0 \rightarrow -\epsilon_0$. The reason lies in the density deviations, which depend on  $U, t,$ and $\epsilon_0$ in different ways. Therefore, the same applies to the limits of what we can consider a weak coupling regime. 

\section{Truncation Effects}\label{sec:app_C}
\renewcommand{\thefigure}{\,C\arabic{figure}}
\setcounter{figure}{0}

The eventual interest of the method relies heavily on the scaling in the amount of measurements required. As mentioned in the text, this is a  criticism common to all strategies based on circuit-knitting techniques.  This amount is  proportional to  the exponentially growing number of  observables,  $\hat E_i$ and $O^{ij}_{L,a}$ with $i  \leq 2^N$.  From the work flow in \figref{fig:workflow} it is apparent that the importance of these quantities in the evaluation of the  cost function $\langle H_{tot}\rangle$  is controlled by the size of the Schmidt coefficients, i.e. Boltzmann weights that are strongly suppressed for excited states. In order to circumvent the exponential slowdown, a truncation of the spectrum needs to be applied, the requirement being that the number of relevant Schmidt coefficients (hence energies below some threshold)  grow polynomially (or at most subexponentially) in the system size. This in turn translates into a scaling of the temperature with the circuit width as we will see.\footnote{The temperature $T$ is not a free parameter if one is interested in the full TFD of a real system in connection with a bath. On the other hand, if one wants to make use of the method to just extract the low energy spectrum, the $T$ can be used as a handle to make the variational obtention well approximated by just a handful of Schmidt coefficients.}
\\

Trying to make  general statements about the impact of such a truncation necessarily confronts with the vast amount of classes to which local Hamiltonians may belong. In order to make some progress we will  consider generic systems exhibiting non-integrable chaotic dynamics. This can have its origin in various mechanisms, like long-range interactions, high dimensionality or periodic driving, among others. 
 In these cases, the many body system is typically compliant with the Eigenstate Thermalisation Hypothesis (see \cite{Dalessio16} for a review).\\

To take a step forward, we start form the observation that, in a generic system with $N$ spins, the number of states $\Omega = 2^N$ grows exponentially with the system size. Making the crude approximation  that $E = N \epsilon$, with $\epsilon$ the average energy per spin, the density of states $\rho(E)$ takes the form

$$
\rho(E) = \frac{d \Omega}{d E} \sim \exp( \alpha E)
$$

with $\alpha = ln 2 / \epsilon$. For chaotic systems, the distribution that governs the spacing of nearby eigenvalues is the Wigner-Dyson distribution which predicts that, for narrow sections of the spectrum, closeby energy eigenstates tend to feel a ``level repulsion''. This, together with the exponential reduction of the spacing, leads to a ``dense'' spectrum which differs qualitatively from integrable systems (which typically follow Poissonian statistics, as a result of eigenstate independence) \cite{Dalessio16}.

In order to show tractability of the method put forth in this work, we now calculate several upper bounds for the temperature  at which diagonalization via EF of a TFD is expected to be of interest. The number of states up to a cutoff energy $E_C$ is given by 
$$
\Omega(E_C) = \int_0^{E_C} \rho(E) dE \sim \frac{\exp( \alpha E_C)}{\alpha}
$$
We would like to set a bound to the scaling of this number in terms of some polynomial  in  the number of states
$$
e^{\alpha E_C} \sim N^p
$$

In this case we get that the cutoff energy scales like $\alpha E_C  \sim  \textrm{p } log(N)$. Setting $E_C \sim N\epsilon \sim N k_B T$ this implies for the 
the cutoff inverse temperature 
\begin{equation}
\beta^{(poly)}_C \sim \frac{\alpha k_B N}{\textrm{p } log(N)}
\label{eq:ndepT}
\end{equation}

This means that for the number of retained states to remain polynomial on the number of spins, the temperature must decrease with the system size with a logarithmic correction. In particular, retaining a fixed number $K$ of excited states ($p=0$) yields  $\beta_C^{K} \sim \alpha k_B N/K$, and the temperature must drop linearly with $N$.

\begin{figure}
\includegraphics[scale=0.45]{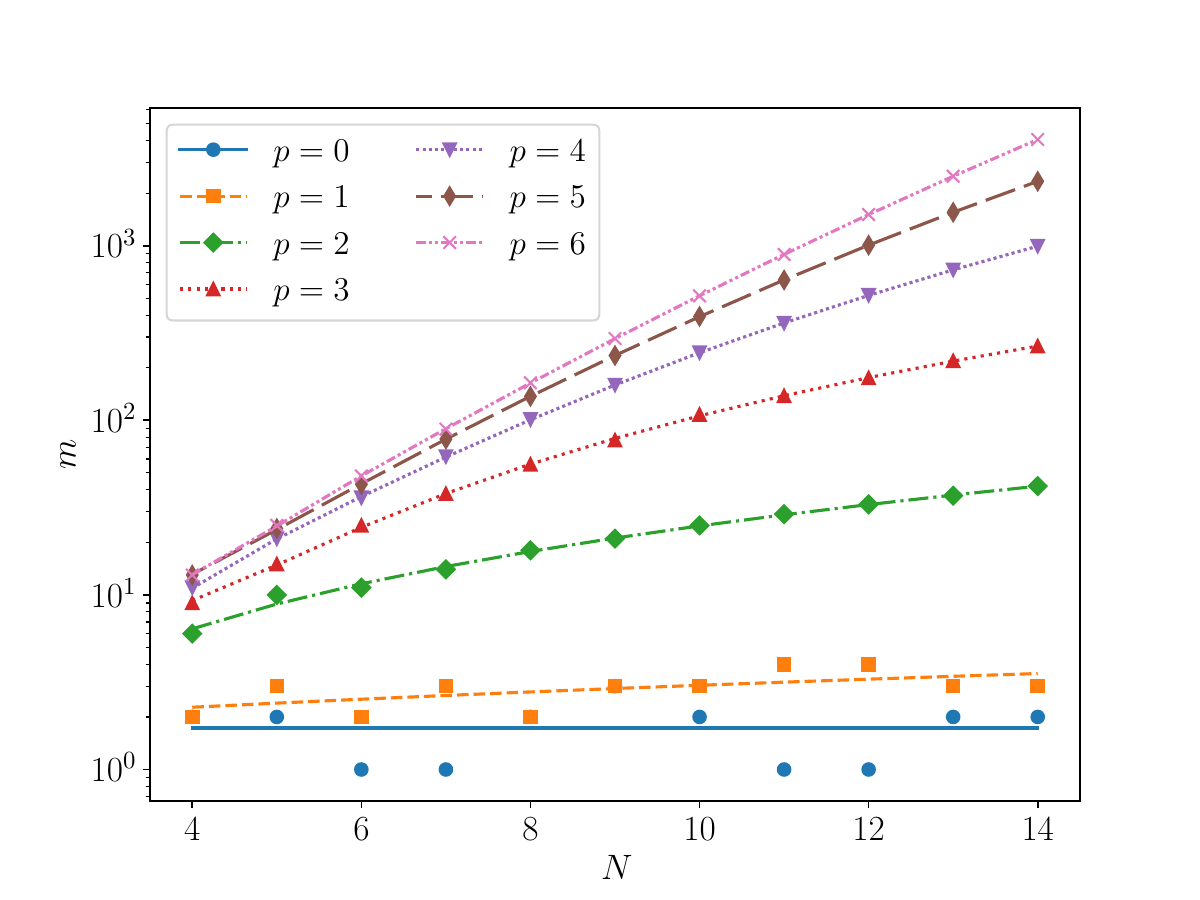}
\caption{Numerical data showing the number of states $m$ kept to get $\left| \left\langle \hbox{TFD}\left(\beta\right)|\tilde{\hbox{TFD}}\left(\beta, m\right)\right\rangle\right|>0.98$ in the Hubbard model with $\beta$ given by \eqref{eq:ndepT}. Dashed lines are fits to polynomials of degree $p$, matching the theoretical expectation exactly.}
\label{Fig:C1}
\end{figure}

We may numerically check \eqref{eq:ndepT} by plotting the overlap of the exact TFD state with its truncated form
\begin{equation}
    \ket{\tilde{\hbox{TFD}}(\beta,m)}=\frac{1}{\sqrt{\sum_k^m e^{-\beta E_k}}}\sum_i^m e^{-\beta E_i/2} \ket{E_i}\otimes \ket{E^*_i},
\end{equation}
In \figref{Fig:C1} we show the value $m$ needed to keep the overlap $\left| \left\langle \hbox{TFD}\left(\beta\right)|\tilde{\hbox{TFD}}\left(\beta, m\right)\right\rangle\right|>0.98$ decreasing the temperature with $N$ as $\beta=\frac{N}{p \log(N)}$, which validates the theoretical expectation given in \eqref{eq:ndepT}.

\begin{figure}
\includegraphics[scale=0.45]{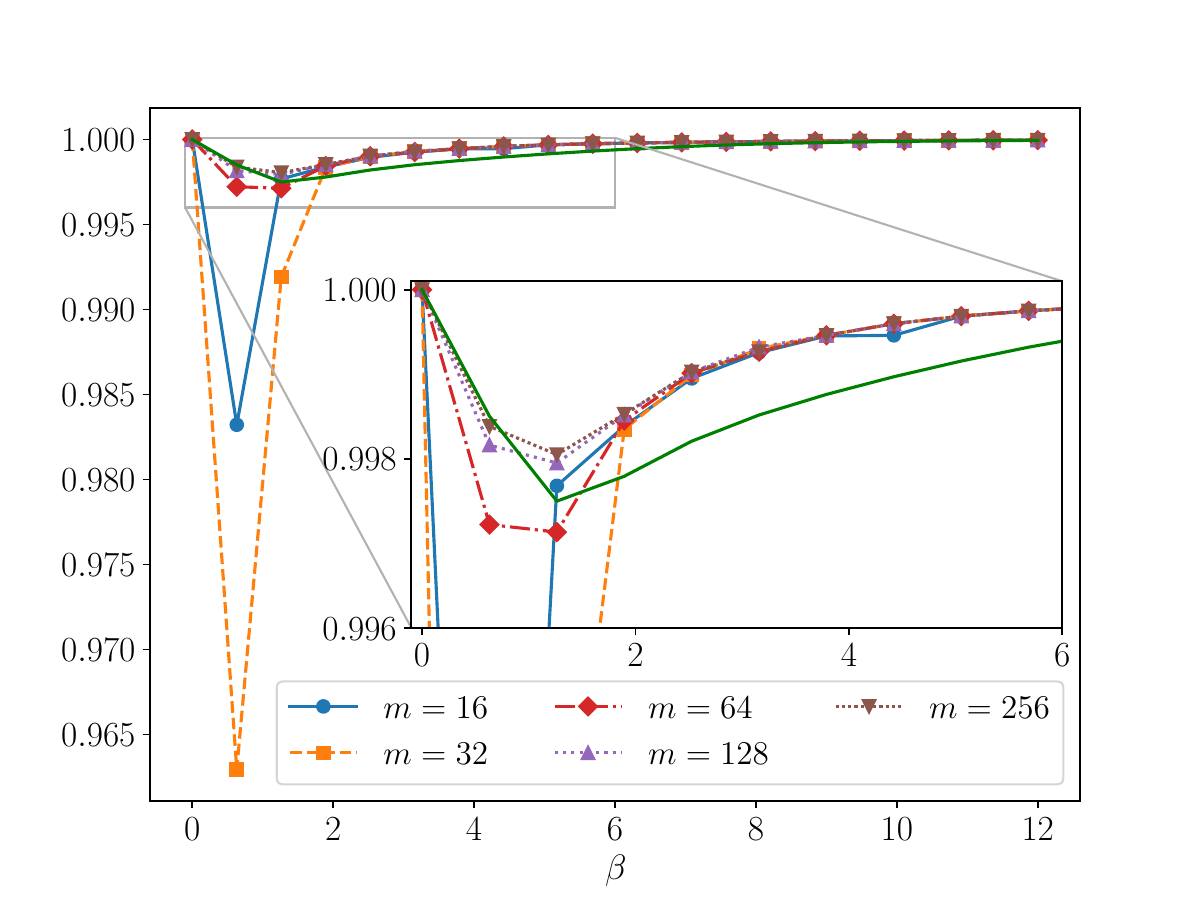}
\caption{Comparison between $\left| \left\langle \hbox{TFD}\left(\beta\right)|\hbox{GS}\left(\beta\right)\right\rangle\right|$ (green solid line) and $\left| \left\langle \hbox{TFD}\left(\beta\right)|\Psi_{opt}\left(\beta, m\right)\right\rangle\right|$ truncating the cost function to $m$ states for $N=8$. The solid line generally bounds from below the optimized results, except for a narrow interval when $m$ is small $(16, 32)$ where the optimization becomes unstable. This shows that the optimization is resilient to truncation.}
\label{Fig:C2}
\end{figure}

In the variational optimization process, truncation is applied not to the variational ansatz \eqref{wave_fun}, but to the number of circuits used to compute the expectation values in \eqref{expectationEF} and \figref{fig:workflow}. This approach preserves the structure of the wave function while focusing the optimization on the low-energy portion of the spectrum. Importantly, the orthogonality of $U\left(\theta\right)$ and the nature of the VQA, which inherently captures the interactions of the Hamiltonian, ensures that the higher part of the spectrum is still reasonably well approximated. As a result, the overlap remains controlled, and the findings in the previous paragraph can be interpreted as a lower bound on the overlap. In \figref{Fig:C2}, we present the results of this truncated optimization, comparing them with the previously discussed overlaps. Our numerical results suggest that the optimization protocol remains robust under truncation, {\em even in the low-$\beta$ regime}. 
Ultimately, this truncation reduces the number of circuits and, thereby, the number of shots. This enhances the feasibility of the protocol in real quantum computer implementations.

\subsection*{Relationship with Iterative Subspace Methods}

The variational diagonalization presented in the main text performs a diagonalization by combining three main ingredients in a variational quantum eigensolver:\\
(i) A TFD ansatz together with an approximate interaction Hamiltonian, which put together allow to find the Schmidt coefficients and the diagonalizing unitary via variational optimisation.\\
(ii) An EF procedure, which halves the Hilbert space of the system being diagonalized. This effectively brings back the task to evaluating  single system, rather than a double system, circuits. \\
(iii) A cutoff temperature $T_C$. Enforcing a temperature cutoff which restricts the diagonalization to the low-energy subspace, as shown above.

This bears some resemblance with iterative subspace methods, which are a class of algorithms aimed at obtaining a subset of $K$ eigenvalues out of the total spectrum of sparse matrices of size $N\gg K$.

In iterative methods, the complexity is reduced from $O(N^3)$ of direct methods to $O(N^2 K)$ by selecting the dimension of the Krylov subspace. At each iteration, these algorithms refine the eigenspace by minimizing the Rayleigh quotient \cite{saad03}. Selecting the cutoff temperature $T_C$ controls the size of the subspace in the diagonalization via EF in a way which is reminiscent of the selection of the number of the subspace dimension $K$.

Dealing with finite precision is a typical bottleneck for iterative subspace methods \cite{abdelfattah21}, which leads to a loss of orthogonality of the internal working basis and to the appearance of spurious eigenvalues which are difficult to distinguish from real ones. Similarly, diagonalization via EF will give wrong results whenever the spacing between eigenvalues cannot be resolved due to a finite amount of measurements. The Wigner-Dyson distribution predicts that eigenstates spacing is dominated by a vanishing exponential \cite{Dalessio16}, which will demand an exponential amount of measurements to retrieve the correct eigenvalue ordering. Therefore, in both cases (classical and quantum), there will be a case-dependent system size in which finite numerical precision (or fluctuations due to a finite amount of measurements) lead to a loss of orthogonality. In both cases either the need for expensive procedures, like re-orthogonalisation (with complexity $O(NK^2)$ \cite{golub13}) or increasing the number of repetitions, will be necessary.

In contrast, in our method, the orthogonality is ensured {\em ab-initio}. In fact, in our approach, even if we truncate the contribution of the exponentially small Boltzmann weights to the evaluation of  the cost function $\langle H_{tot}\rangle$, the protocol always yields a unitary matrix that rotates the computational basis to an approximate (yet  orthonormal) energy basis. 
 In a sense, our algorithm is closer in spirit  to the family of subspace-search variational algorithms \cite{Nakanishi_2019, Benavides-Riveros:2022two, Hong:2023atp} which are, by construction, orthogonality preserving. 
 
\section{Boltzmann distribution loading}\label{sec:app_D}
\renewcommand{\thefigure}{\,D\arabic{figure}}
\setcounter{figure}{0}
\begin{figure*}[htbp!]
\includegraphics[scale=0.5]{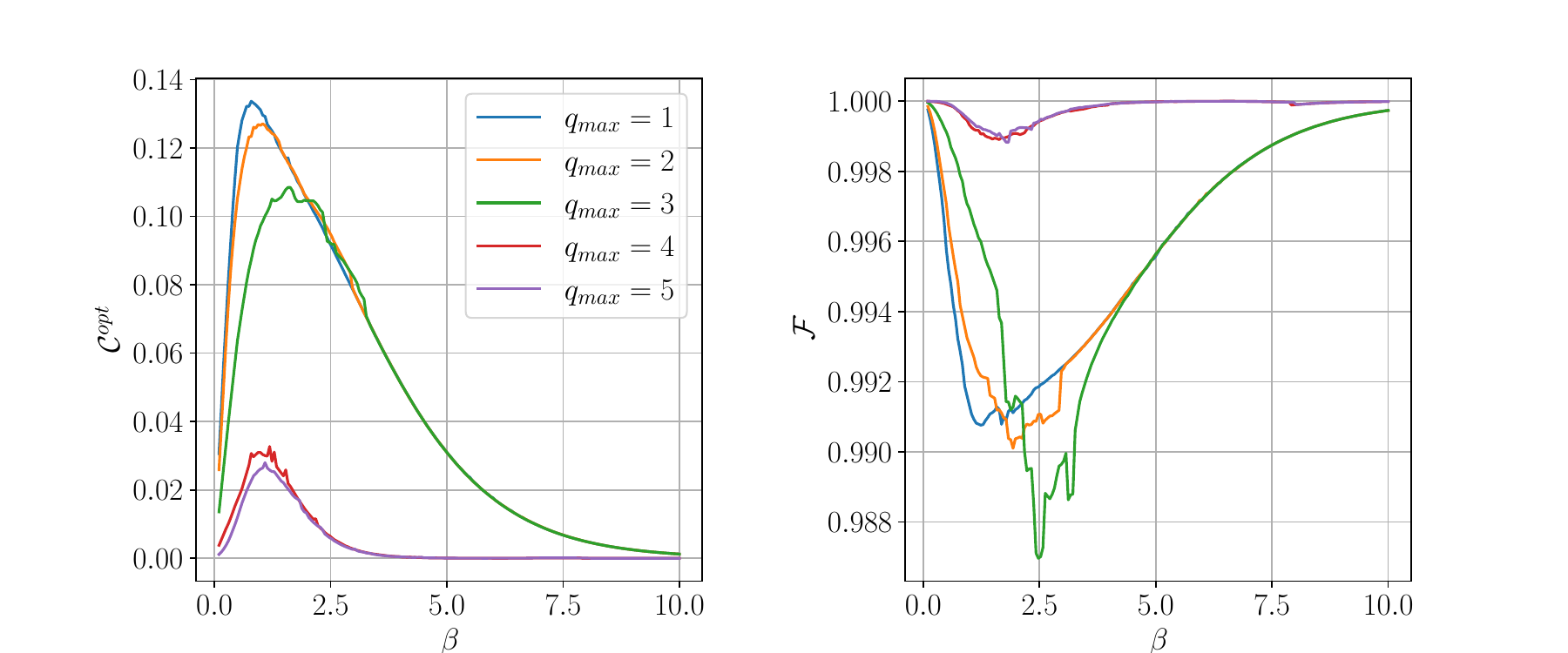}
\caption{Results of the optimization of the loading of Boltzmann coefficients. Left, the value of the cost function after the optimization. Right, the overlap with the target state. At $q_{max}>3$, there is a noticeable jump that significantly improves the results.}
\label{Fig:D1}
\end{figure*}
As mentioned in Sec. \ref{sec:forging} of the main text, if the optimization in \figref{fig:workflow} succeeds, the outputs of the EF protocol yield the variational estimations of the energies $\tilde{E}_i\left(\btheta_{opt}\right) \approx E_i$ and the  change of basis to the energy eigenbasis, $U(\btheta_{opt})$. If one is interested in preparing the  TFD itself it is necessary to implement a unitary $U_{\lambda}$ that loads the Boltzmann distribution given by $\tilde{\lambda}_i = e^{-\beta \tilde E_i/2}/\sqrt{Z}$
\begin{equation}    U_{\lambda}\ket{0}=\sum_i\tilde{\lambda}_i \ket{b_i}. \label{D1}
\end{equation}
In this appendix we will discuss how to perform this task variationally. In the same spirit as in the previous part, the free fermion case admits an exact solution, and we will use it as a {\em warm start} for another variational circuit that tackles the interacting case.

Indeed, in the free fermion case
$$H = \sum_k \omega_k a_k^\dagger a_k$$
we can find $U_\lambda$ exactly, given the fact that we know the exact expression for the TFD as given in \eqref{zerobeta1}.  The circuit that generates it, disregarding again the minus signs that come from the fermionic statistics,
only requires $N$ $R_y$ gates and $N$ CNOT gates.
Namely, using that 
$$
R_y(2\varphi_k)\ket{0} = \cos(\varphi_k)\ket{0}+\sin(\varphi_k)\ket{1} = u_k\ket{0}+v_k\ket{1}.
$$
again with $\varphi_k = \arctan e^{-\beta \omega_k/2} $, the circuit that gets the job done is the following. 

\begin{center}
\begin{quantikz}[row sep=0.1cm,column sep=0.5cm]
&\gate{R_Y(2\varphi_0)}& \ctrl{3} & \qw & \qw  & \qw \\
&\gate{R_Y(2\varphi_1)}& \qw &\ctrl{3} & \qw & \qw\\
&\gate{R_Y(2\varphi_2)}& \qw & \qw &\ctrl{3} & \qw \\
&\qw&\targ{}&\qw&\qw& \qw   \\
&\qw&\qw&\targ{}&\qw& \qw \\
&\qw&\qw&\qw&\targ{}& \qw
\end{quantikz}
\end{center}

Where we have used $N=3$ for concreteness but can be directly generalized to arbitrary $N$. Thus, in the free fermion case, the exact expression for the loading circuit $U_\lambda$ is
\begin{equation}
    U_\lambda = \bigotimes_i^N R_y\left(2\varphi_i\left(\omega_i\right)\right) \, .
\end{equation}
In the general interacting case, $U_\lambda$ will have to introduce entanglement, but its closed form is a priori not known. Given that we are in a perturbative regime around the free fermion scenario, we suggest to address this problem variationally with the following ansatz
\begin{equation}
U_{\lambda}=\tilde{U}_\lambda(\bphi)\bigotimes_i^N R_y\left(2\varphi_i\left(\tilde{\omega}_i\right)\right),\label{utilde}
\end{equation}
where $\tilde{U}_\lambda(\bphi)$ is a real variational ansatz that depends on some set of parameters $\bphi$ and has no effect when they are set to zero. This is required so that the beginning of the optimization is a warm start solution that prepares the Boltzmann distribution of $\tilde{\omega}_i$ (corresponding to the optimized quadratic contribution \eqref{wshift}). 

In order to optimize $\tilde U_\lambda(\bphi)$ we  propose the cost function 
\begin{equation*}
    \mathcal{C}(\bphi) = \sum_{i=1}^{2^N} \left|\lambda_i^2 - p_i(\bvarphi, \bphi)\right|\, ,
\end{equation*}
where $\lambda_i$ are the data, and $p_i(\bvarphi, \bphi)$ are the measured probabilities of obtaining each of the states in the computational basis of a given variational circuit. Notice that this cost function is not sensitive to local signs for the amplitudes $c_i = \pm\sqrt{p_i}$. However, given that $U_\lambda(\bvarphi, \bphi)$ is real and warm started, it is expected that the amplitudes will remain positive, rather than jump over to the opposite sign.

In \figref{Fig:D1}, we show the results of this loading process for the variational energies obtained in \figref{fig:simulations}(e).  We also plot the overlap $\mathcal{F}=\sum_i \tilde{\lambda}_i\left| \bra{b_i}U_\lambda \left(\bvarphi, \bphi\right)\ket{0}\right|$ between the obtained state and the exact target state \eqref{D1}. As for the variational ansatz, we used
\begin{equation}
    \tilde{U}_\lambda (\bphi) = \prod_l^\mathcal{L}\prod_{q=1}^{q_{max}} \prod_{i=0}^{N-q-1} e^{-i \phi_{q,l} Z_i Y_{i+q}}\, .
\end{equation}
The first product refers to the layers and $q_{max}\leq N$ is an upper bound to $q$.
The terms of the third product are arranged to minimize the circuit depth. The parameters $\bvarphi$ are also left free during the optimization to increase expressivity. The total number of parameters to be optimized is $N + \mathcal{L}q_{max}$. This is $N$ parameters $\bvarphi$ and $\mathcal{L}q_{max}$ parameters $\bphi_{q,l}$. We find experimentally that it is sufficient to consider $\mathcal{L}=1$.

We see that when increasing $q_{max}$ from $3$ to $4$, the results improve significantly, achieving overlaps $>0.999$ for every $\beta$, which is effectively a perfect optimization.

%

\end{document}